\documentclass[pre,twocolumn]{revtex4}
\usepackage{epsfig}
\usepackage{bm}
\usepackage{amsmath}

\begin{document}

\title{Chaotic quasi-point vortices and inertial range in classic and quantum two-dimensional turbulence}

\author{A. Bershadskii}

\affiliation{
ICAR, P.O. Box 31155, Jerusalem 91000, Israel
}

\begin{abstract}
  Effects of quasi-point vortices on the inertial range of scales in homogeneous two-dimensional turbulence (classic and quantum) have been studied using the notion of distributed chaos. Results of direct numerical simulations of decaying turbulence, turbulence with small-scale forcing, and turbulent thermal convection on a sphere as well as results of the Global Atmospheric Sampling Program (GASP) measurements in the tropospheric and stratospheric turbulence over mountainous terrain (the small-scale forcing) have been used for this purpose. The superfluid and Bose-Einstein two-dimensional turbulence have been discussed in this context using the results of direct numerical simulations of the HVBK and Gross-Pitaevskii models, and the laboratory experiments. The Ginsburg-Landau model was also briefly discussed. 
\end{abstract}

\maketitle

\section{Quasi-point vortices and inertial range}
 
 Let us start from freely decaying homogeneous isotropic two-dimensional turbulence. More precisely we will be interested in the stage of decay when a system of well-separated quasi-point vortices has been developed. This system contains main part (almost entire) of the flow enstrophy \cite{br}-\cite{w1}. Similar phenomenon is also typical for homogeneous isotropic two-dimensional turbulence with small-scale forcing \cite{sy2},\cite{sy1}. This case will be considered below for direct numerical simulations and for atmospheric turbulence where the large aspect ratio of lateral to vertical length scales allows quasi-two-dimensional approximation (though one should take into account that properties of quasi-two-dimensional turbulence can be very different from those of strictly two-dimensional one \cite{btk}) . 
 
  While the quasi-point vortices are not too close to each other they can be considered as an adiabatically changing with time Hamiltonian system \cite{es} and described by corresponding equations
$$
\Gamma_i \dot{x_i} = \frac{\partial H}{\partial y_i}, ~~~~~~~~~~  \Gamma_i \dot{y_i} = -\frac{\partial H}{\partial x_i} 
  \eqno{(1)}
$$
with the Hamiltonian
$$
H = -\frac{1}{4\pi} \sum_{i<j}^N \Gamma_i\Gamma_j \ln [(x_i-x_j)^2 +(y_i-y_j)^2],   \eqno{(2)}
$$
where $N$ is the number of the quasi-point vortices,  $\Gamma_i$ are their strengths (circulations), $x_i$ and $y_i$ are coordinates of their effective centres on the plane (see, for instance, Refs. \cite{aref},\cite{y1}). 

  The Hamiltonian system  has three formal invariants on the unbounded plane 
$$
P_x = \sum_i \Gamma_i x_i,~~~~~~~~~~~ P_y = \sum_i \Gamma_i y_i,  \eqno{(3)}
$$
and 
$$
\mathcal{I}= \sum_i \Gamma_i (x_i^2+y_i^2)    \eqno{(4)}
$$

The invariants Eq. (3) are the two components of the linear momentum ${\bf P}$ and the invariant $\mathcal{I}$ Eq. (4) is angular momentum for the Hamiltonian system Eqs. (1-2). The Noether's theorem relates conservation of ${\bf P}$ and $\mathcal{I}$ to the spatial homogeneity (translational symmetry) and to spatial isotropy (rotational symmetry) correspondingly (see, for instance, Ref. \cite{aref} for the classic case and Ref. \cite{luc} for the quantum case).  At certain conditions these (formal) invariants can be considered as adiabatic invariants for the inertial range of scales in addition to the Kolmogorov-Obukhov adiabatic invariant $\varepsilon$ - energy dissipation rate \cite{my}. For the two-dimensional turbulence additional support for the adiabatic invariance of $\varepsilon$ in the inertial range of scales is provided by the conservation of the enstrophy $\Omega$ for ideal incompressible fluid motions and the relationship
$$
\varepsilon = \nu~ \Omega,  \eqno{(5)}
$$   
where $\nu$ is viscosity.

Actually, we will be interested in two combined adiabatic invariants:
$$
I_1 = ||{\bf P}||\varepsilon   \eqno{(6)}
$$
and
$$
I_2 = |\mathcal{I}| \varepsilon \eqno{(7)}  
$$
taking into account both the quasi-point vortices and the $\varepsilon$-phenomenology (not always related to the cascade mechanisms \cite{b1}).

\section{Distributed chaos approach to the inertial range} 

At the onset of isotropic homogeneous turbulence spectral decay of kinetic energy has exponential form \cite{kds}
$$
E(k) \propto \exp-(k/k_c)  \eqno{(8)}
$$    

  When the turbulence is developing the parameter $k_c$ becomes fluctuating and ensemble average is needed in order to calculate the kinetic energy spectrum 
$$
E(k) \propto \int_0^{\infty} P(k_c) \exp -(k/k_c)dk_c  \propto \exp-(k/k_{\beta})^{\beta}  \eqno{(9)}
$$    
that uses probability distribution $P(k_c)$. A generalization of the exponential spectrum Eq. (8) to the stretched exponential form Eq. (9) can be also used. From the Eq. (9) one can estimate the asymptote of the $P(k_c)$ at large $k_c$ \cite{jon}
$$
P(k_c) \propto k_c^{-1 + \beta/[2(1-\beta)]}~\exp(-bk_c^{\beta/(1-\beta)}) \eqno{(10)}
$$

  On the other hand the asymptote of $P(k_c)$ can be estimated from a physical consideration. Let as assume that there exists an asymptotical relationship between the characteristic velocity $v_c$ and the $k_c$ dominated by one of the adiabatic invariants Eqs. (6-7). Then from the dimensional considerations we obtain
$$
v_c \propto I_1^{1/4}~k_c^{1/4}  \eqno{(11)}
$$  
or
$$
v_c \propto I_2^{1/4}~k_c^{1/2}.  \eqno{(12)}
$$   
In a general form
$$
v_c \propto k_c^{\alpha}  \eqno{(13)}
$$   
Then for Gaussian distribution of the characteristic velocity $v_c$ we obtain from the Eqs. (10) and (13)
$$
\beta = \frac{2\alpha}{1+2\alpha}   \eqno{(14)}
$$
That results in
$$
E(k) \propto \exp-(k/k_{\beta})^{1/3}  \eqno{(15)}
$$
for the Eq. (11) (domination of the spatial homogeneity over the inertial range), or in 
$$
E(k) \propto \exp-(k/k_{\beta})^{1/2}  \eqno{(16)}
$$
for the Eq. (12) (i.e. domination of the spatial isotropy over the inertial range).\\
%%%%%%%%%%%%%%% 1 %%%%%%%%%%%%%%%%%%
\begin{figure} \vspace{-1.2cm}\centering
\epsfig{width=.45 \textwidth,file=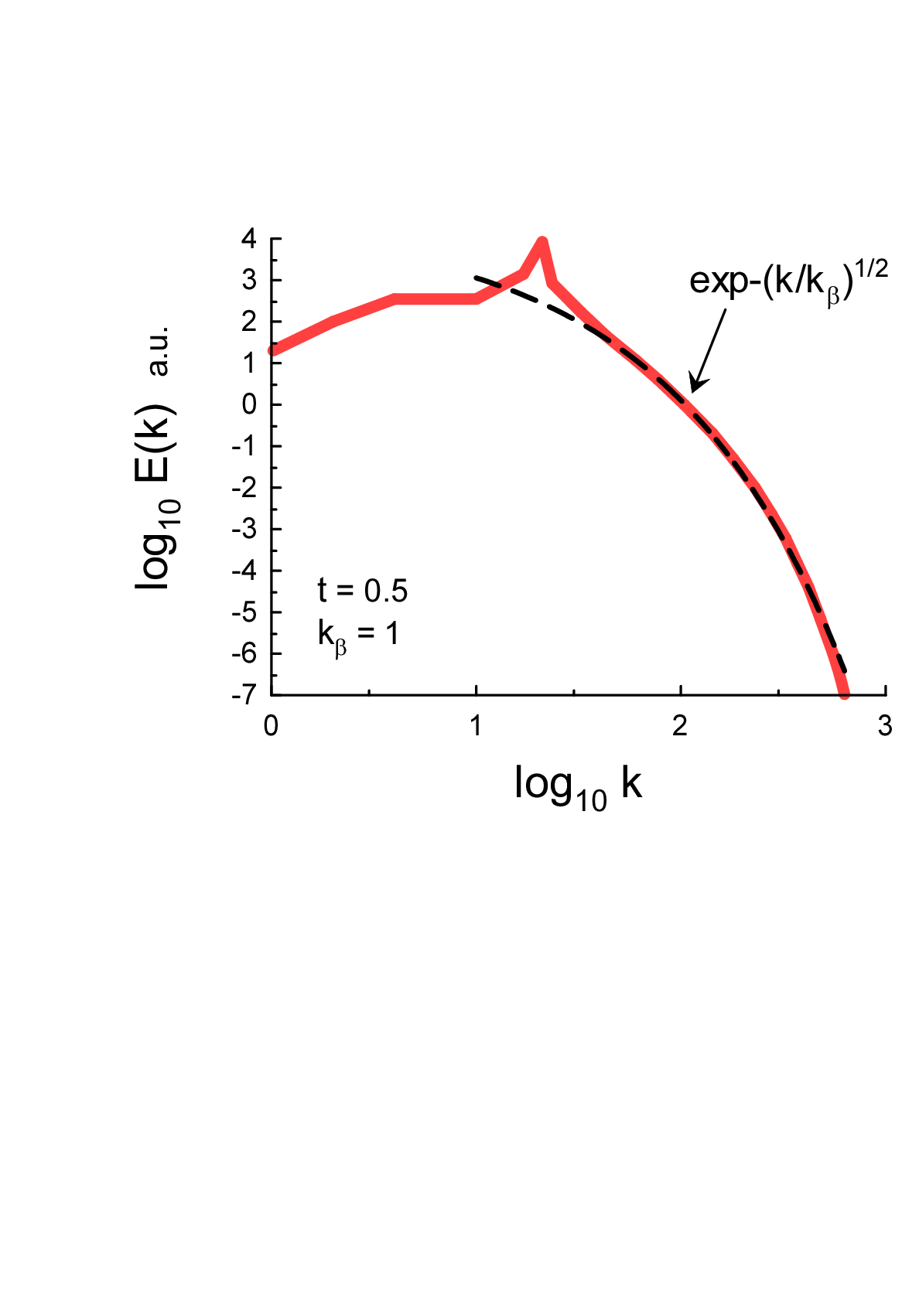} \vspace{-4.5cm}
\caption{Kinetic energy spectrum for decaying homogeneous isotropic two-dimensional turbulence at $t=0.5$ (the DNS time scales).} 
\end{figure}
%%%%%%%%%%%%%%%%%%%%%%%%%%%%%%%%%%% 
%%%%%%%%%%%%%%% 2 %%%%%%%%%%%%%%%%%%
\begin{figure} \vspace{-0.5cm}\centering
\epsfig{width=.45\textwidth,file=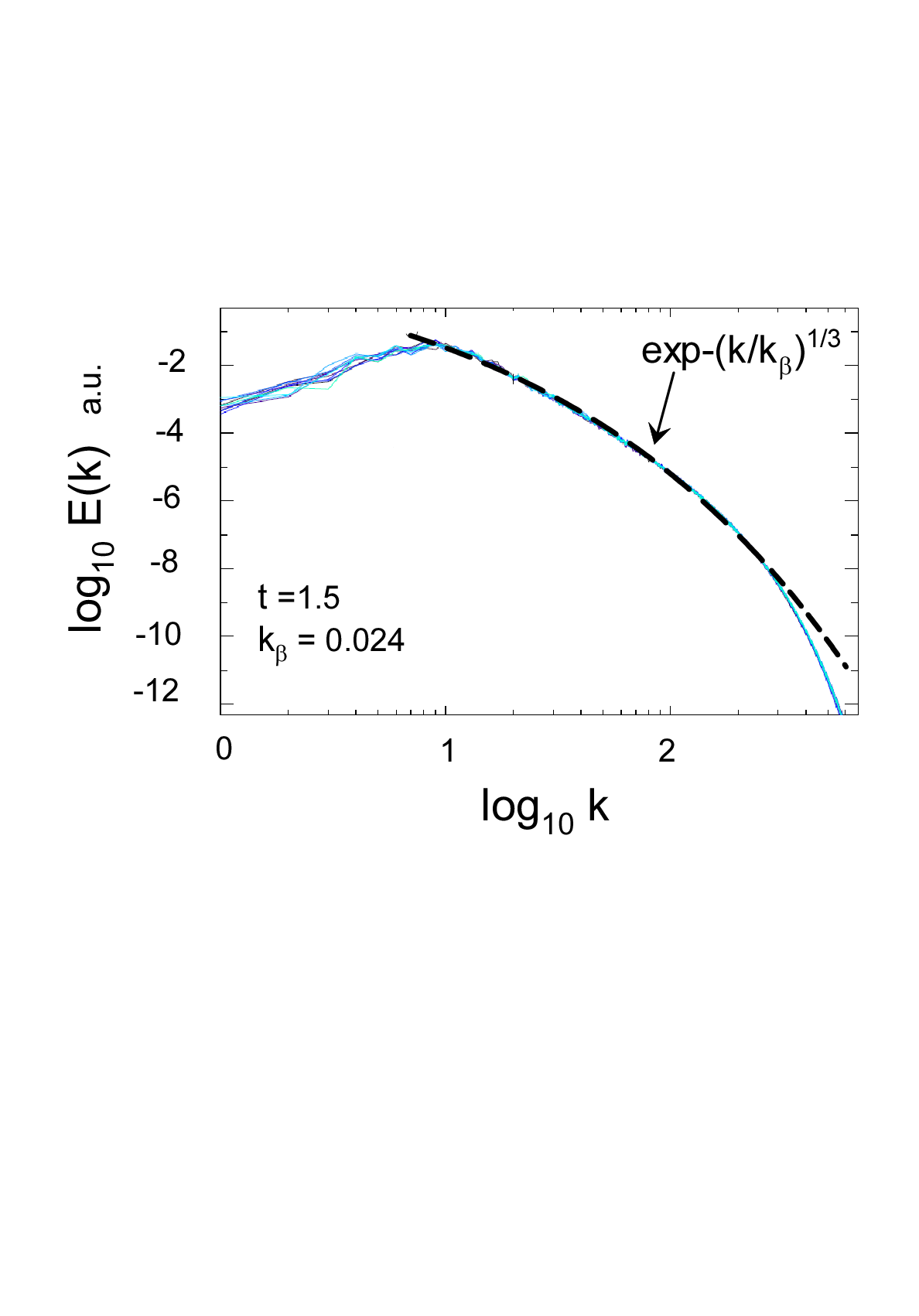} \vspace{-4.2cm}
\caption{Kinetic energy spectrum for decaying homogeneous isotropic two-dimensional turbulence at $t=1.5$ (ten randomly picked realizations, without averaging). } 
\end{figure}
%%%%%%%%%%%%%%%%%%%%%%%%%%%%%%%%%%%    

\section{Direct numerical simulations of decaying turbulence}

  For decaying three-dimensional isotropic homogeneous turbulence the problem of competition between domination of the spatial isotropy (Loitsyanskii invariant \cite{my}) and of the spatial homogeneity (Birkhoff-Saffman invariant \cite{saf}) is a long standing one. Decay of the integral characteristics of the flow can also depend on the way the turbulence is generated (see, for instance, Ref. \cite{vv} and references therein). Analogous situation takes also place for decay of the integral characteristics of two-dimensional turbulence (see, for instance, Ref. \cite{y1} and references therein). \\
  
    It is shown in Ref. \cite{b2} that for three-dimensional isotropic homogeneous turbulence isotropy dominated and homogeneity dominated attractors (i- and h-attractor respectively) have different basins of attraction. The i-basin of attraction (set of the initial conditions resulting in the i-attractor) is thin and small in comparison with the h-basin (set of the initial conditions resulting in the h-attractor). Therefore, in the decaying turbulence the i-attractor approaches its fully developed state earlier than the h-attractor. However at a more advance stage of the decay the h-attractor takes its proper domination due to its larger basin of attraction. The same consideration can be also applied to the inertial range of the decaying two-dimensional turbulence (replacing the Birkhoff-Saffman and Loitsyanskii invariants by the $I_1$ and $I_2$ invariants correspondingly). 
     
%%%%%%%%%%%%%%% 3 %%%%%%%%%%%%%%%%%%
\begin{figure} \vspace{-1.5cm}\centering
\epsfig{width=.45\textwidth,file=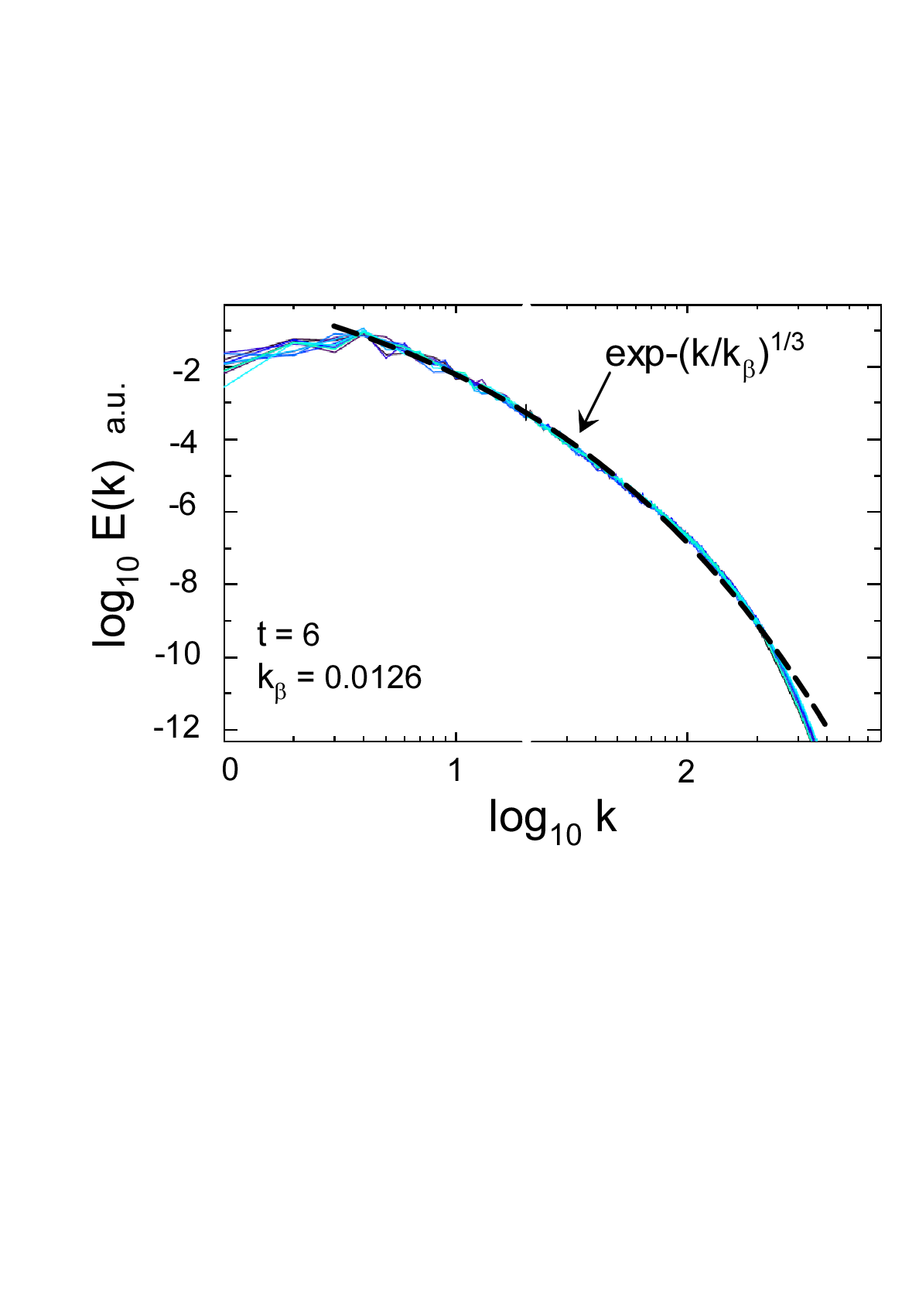} \vspace{-4.2cm}
\caption{As in Fig. 2 but for t=6.} 
\end{figure}
%%%%%%%%%%%%%%%%%%%%%%%%%%%%%%%%%%%  
  %%%%%%%%%%%%%%% 4%%%%%%%%%%%%%%%%%%
\begin{figure} \vspace{-0.5cm}\centering
\epsfig{width=.45\textwidth,file=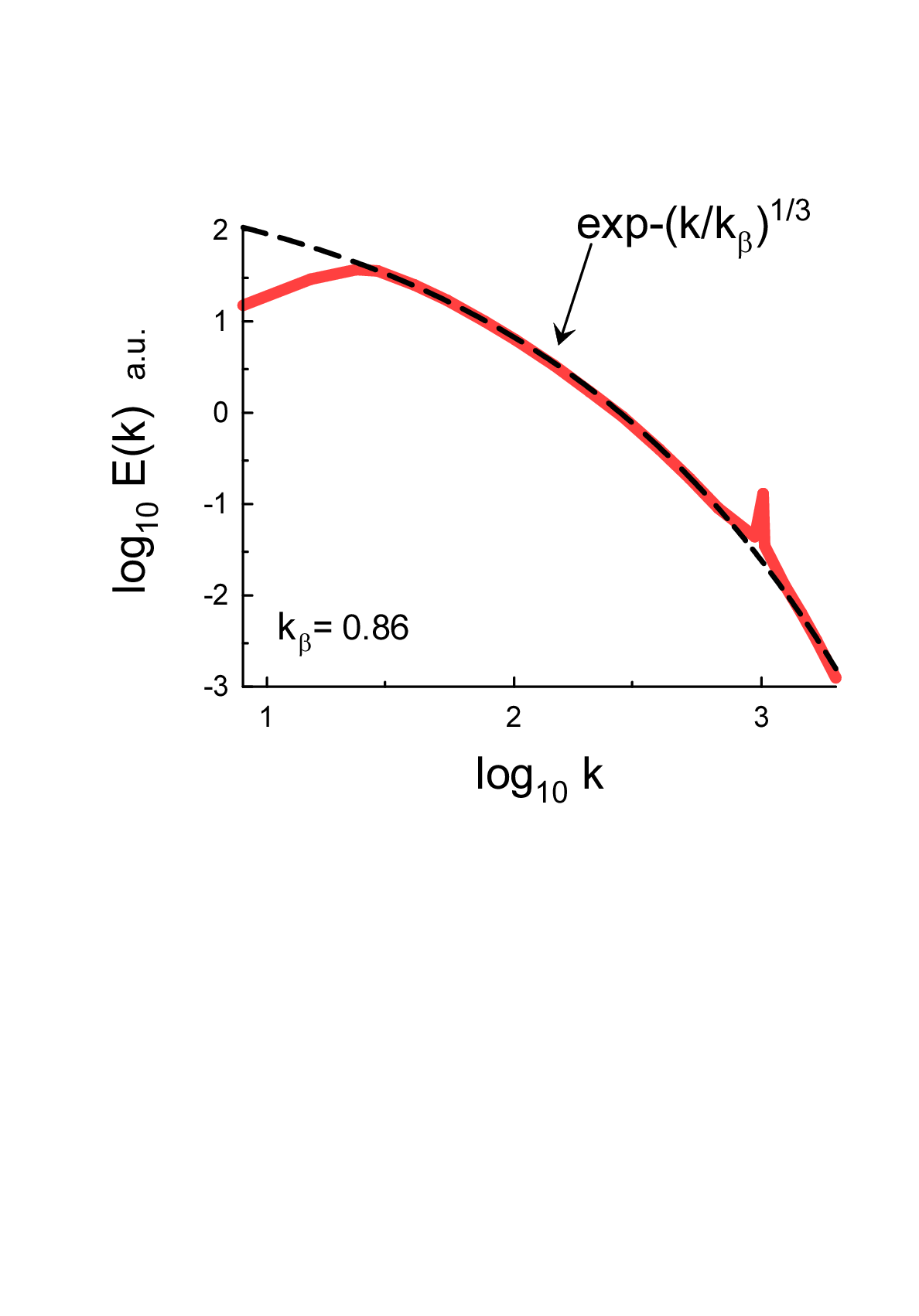} \vspace{-4.4cm}
\caption{Kinetic energy spectrum for the small-scale forcing turbulence.} 
\end{figure}
%%%%%%%%%%%%%%%%%%%%%%%%%%%%%%%%%%%%    

   Figure 1 shows (in the log-log scales) kinetic energy spectrum obtained in direct numerical simulation (DNS) of freely decaying homogeneous isotropic turbulence reported in Ref. \cite{mp} (the spectral data were taken from Fig. 1a of the Ref. \cite{mp} for $t=0.5$ in the DNS time scales). Random superpositions of harmonic modes with wave numbers between k = 18 and 22 were taken as initial conditions for this DNS with periodic boundary conditions. The dashed curve corresponds to the stretched exponential spectrum Eq. (16) (i.e. to the isotropy dominated distributed chaos as it was discussed above), while figure 2 shows kinetic energy spectrum obtained in the DNS for $t=1.5$. The spectral data for the Fig. 2 were taken from Fig. 4 of the Ref. \cite{mp} (ten randomly picked realizations with different random initial phases, without averaging). The dashed curve in the Fig. 2 corresponds to the stretched exponential spectrum Eq. (15), i.e. to the homogeneity dominated distributed chaos as it was discussed above. Analogous spectrum has been shown in figure 3 for $t=6$ (maximal time of the DNS computations). 

   It is interesting to compare these results with the power spectra of passive scalar mixing by chaotic motions of the quasi-point vortices Ref. \cite{b1}.

\section{Small-scale forcing}

 It was already mentioned in Section I that the system of well-separated quasi-point vortices is also developing in homogeneous isotropic two-dimensional turbulence with small-scale forcing \cite{sy2},\cite{sy1}. In recent Ref. \cite{bs} results of direct numerical simulations with a narrow-band forcing with the characteristic forcing wavenumber $k_f = 1024 $ (see a peak in the kinetic energy spectrum shown in Fig. 4) were reported. The spectral data for the Fig. 4 were taken from Fig. 8 of the Ref. \cite{bs}. The forcing was delta-correlated in time and a hyperviscosity was applied. The initial and boundary conditions were a state of no flow and periodic respectively. The dashed curve in Fig. 4 corresponds to the stretched exponential spectrum Eq. (15), i.e. to the homogeneity-dominated distributed chaos (cf. Figs. 2,3). 
  
  It should be noted that using their observations the authors of the Ref. \cite{bs} concluded that the energy spectrum evolves adiabatically in the inertial range of scales and "due solely to the evolution" of the quasi-point vortices population.
  
\section{Thermal (Rayleigh-B\'{e}nard) convection on a hemisphere}  

  On a sphere Hamiltonian for the system of the point vortices can be written as \cite{kur}
$$
H = -\frac{1}{4\pi R^2} \sum_{1\leq j < i \leq N} \Gamma_j\Gamma_i \ln [2R^2(1 - \cos \gamma_{ij})]  \eqno{(17)}
$$    
where $R$ is radius of the sphere and $\gamma_{ij}$ are the angles between the radii
vectors of the $i$ and $j$ point vortices with respect to the sphere center. Due to the
Noether's theorem the rotational symmetry of the system provides three invariants
$$
M_1 = R\sum_i \Gamma_i \sin \theta_i \cos \phi_i \eqno{(18)}
$$
$$
M_2 = R\sum_i \Gamma_i \sin \theta_i \sin \phi_i \eqno{(19)}
$$  
$$
M_3 = R\sum_i \Gamma_i \cos \theta_i  \eqno{(20)}
$$  
where $\theta_i$ and $\phi_i$ are the spherical coordinates of the $i$th vortex.
These invariants commute as components of angular momentum $||{\bf M}||$. From the dimensional considerations we obtain
$$
v_c \propto (||{\bf M||\varepsilon})^{1/4}~k_c^{1/4}  \eqno{(21)}
$$  
and from the Eqs. (13-14) the power spectrum Eq. (15) for the distributed chaos.\\

Results of a DNS of a thermal (Rayleigh-B\'{e}nard) convection on a hemisphere were reported in recent Ref. \cite{brun}. In this DNS a gradient of temperature between the heated equator and the pole creates a thermal convection with Rayleigh number $Ra =10^{10}$ and Prandtl number $Pr= 7$.\\

  Figure 5 shows kinetic energy spectrum obtained in this DNS (the spectral data were taken from Fig. 18 of the Ref. \cite{brun} and corresponds to a stationary state). The variable $k$ in this figure is the spherical wavenumber. The dashed curve corresponds to the stretched exponential spectrum Eq. (15).

%%%%%%%%%%%%%%% 5 %%%%%%%%%%%%%%%%%%
\begin{figure} \vspace{-1.3cm}\centering
\epsfig{width=.45\textwidth,file=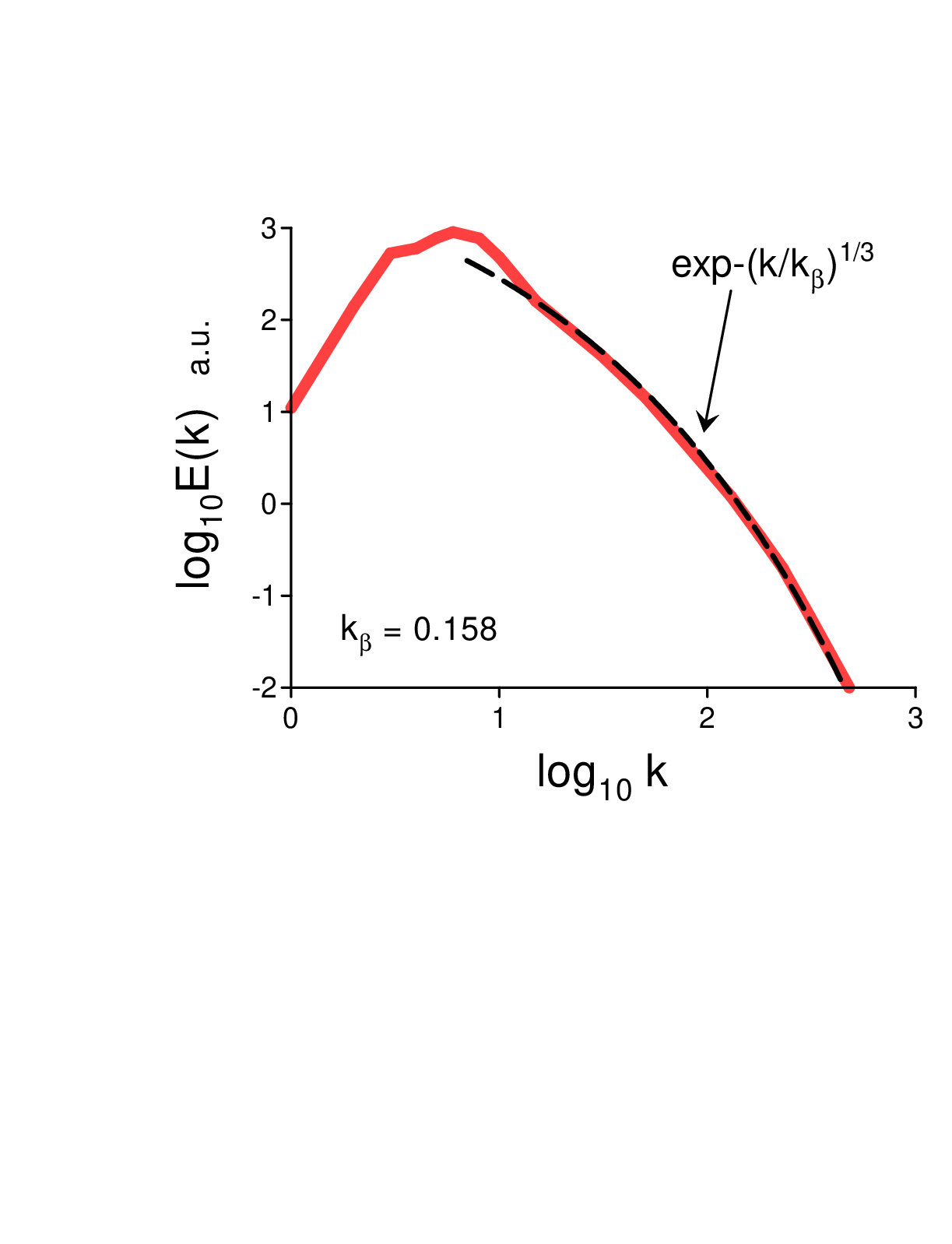} \vspace{-3.7cm}
\caption{Kinetic energy  spectrum for a thermal (Rayleigh-B\'{e}nard) convection on a hemisphere. } 
\end{figure}
%%%%%%%%%%%%%%%%%%%%%%%%%%%%%%%%%%%   
  
%%%%%%%%%%%%%%% 6 %%%%%%%%%%%%%%%%%%
\begin{figure} \vspace{-0.1cm}\centering
\epsfig{width=.45\textwidth,file=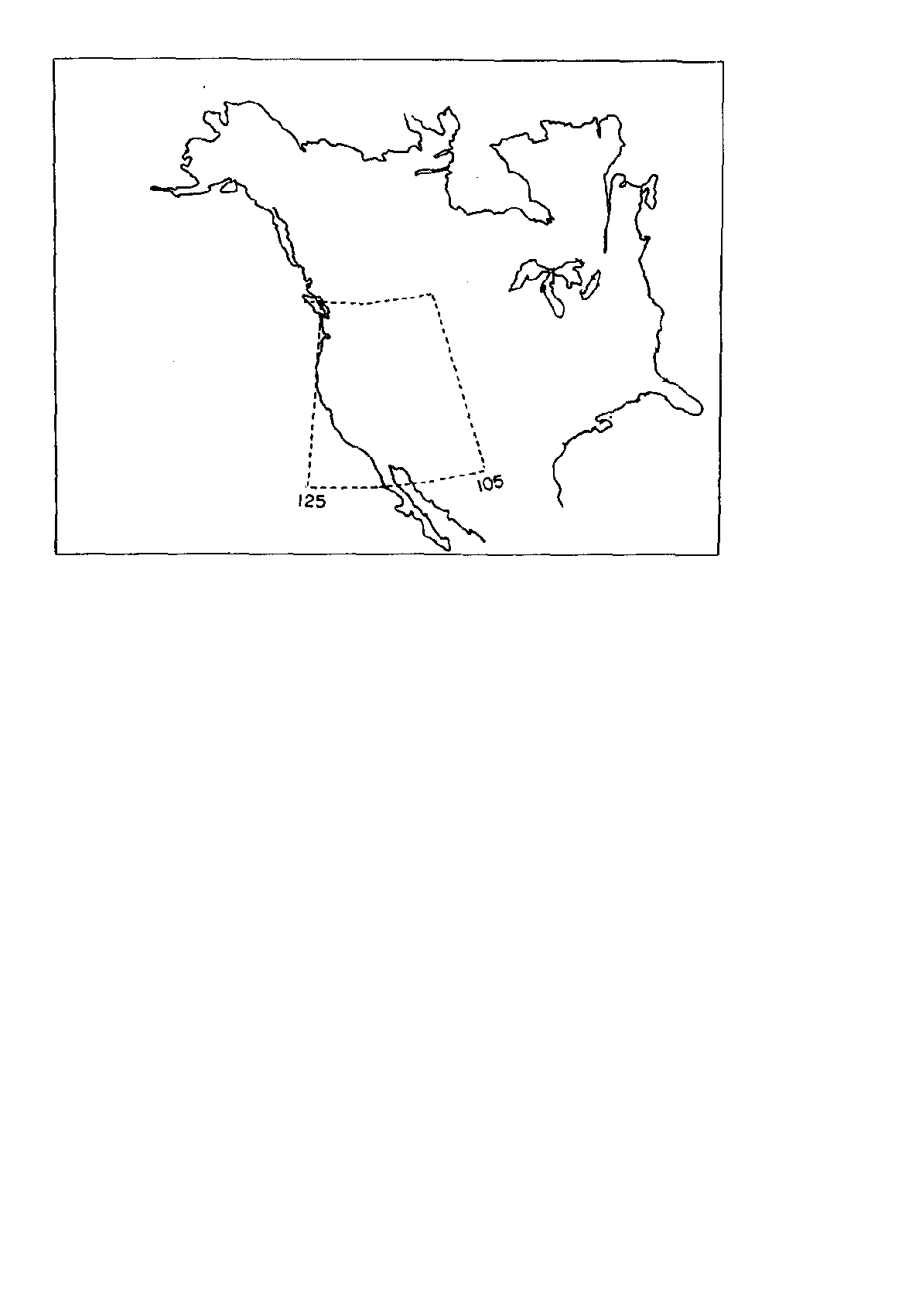} \vspace{-6cm}
\caption{Location of the GASP measurements area over the western U.S.A. (the region bounded by the dashed curves). } 
\end{figure}
%%%%%%%%%%%%%%%%%%%%%%%%%%%%%%%%%%%%
%%%%%%%%%%%%%%% 7 %%%%%%%%%%%%%%%%%%
\begin{figure} \vspace{-0.5cm}\centering
\epsfig{width=.45\textwidth,file=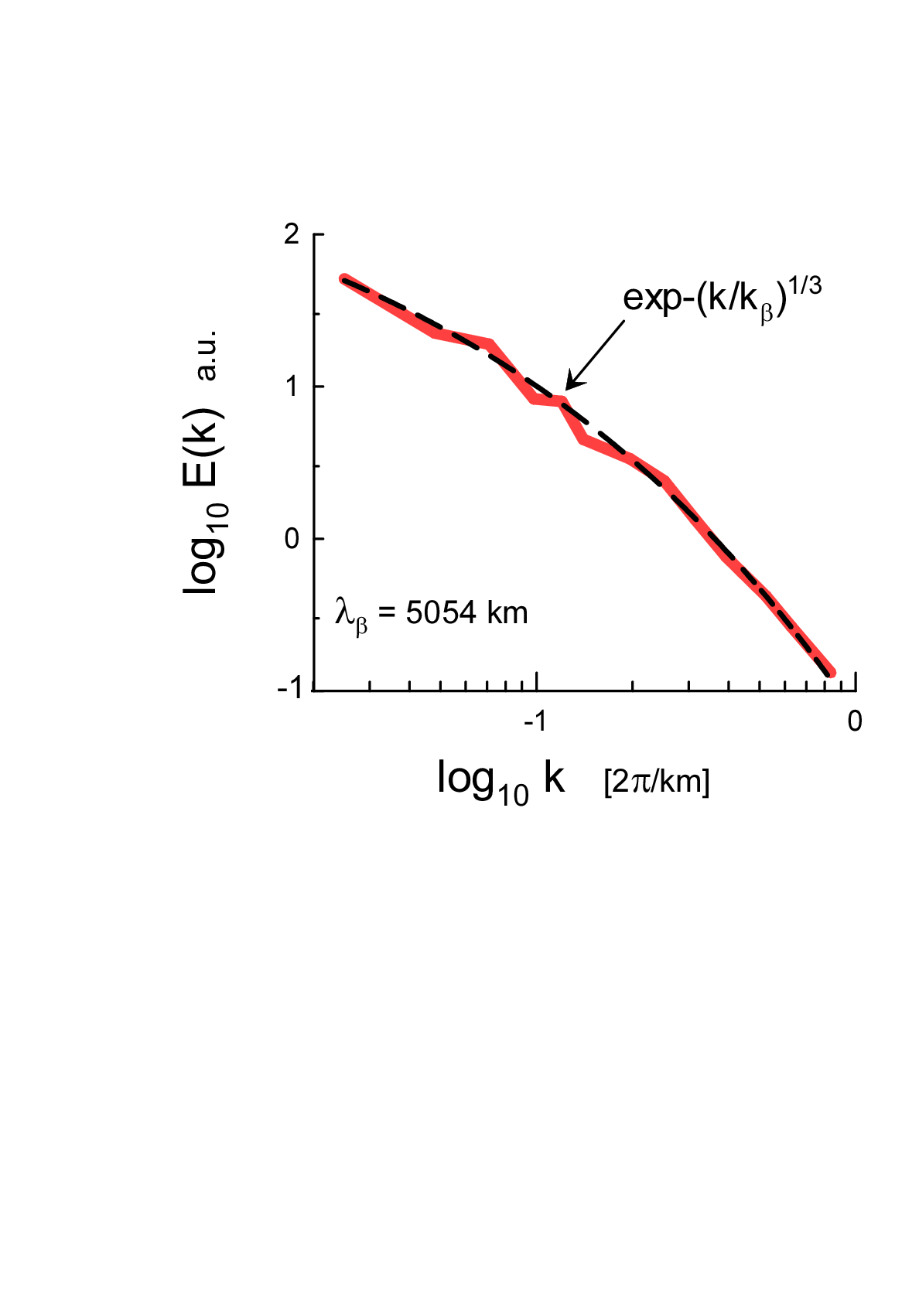} \vspace{-4.4cm}
\caption{Average power spectrum of zonal wind speed for high wind speed cases (troposphere).  } 
\end{figure}
%%%%%%%%%%%%%%%%%%%%%%%%%%%%%%%%%%%%   

\section{Atmospheric turbulence over rough terrain}

   The atmospheric variability depends on the underlying terrain. The data obtained by the seminal Global Atmospheric Sampling Program (GASP) measurements (using commercial aircraft) \cite{nfg} shows that kinetic energy variances can be about six times larger over rough (mountainous) terrain than over plains and ocean. A significant difference has been also observed in the scales of the forcing. The authors of the Ref. \cite{nfg} emphasize the role of the {\it small-scale} forcing in quasi-two-dimensional turbulence over the rough terrain of the western U.S.A. (see Fig. 6 adapted from the Ref. \cite{nfg}) and estimate the range of scales for this forcing as 1-3 km. Since we already know that the inertial range in stationary homogeneous two-dimensional turbulence with small-scale forcing is under the strong effect of the quasi-point vortices (see Fig. 4) it is interesting to look at the kinetic energy spectra reported in the Ref. \cite{nfg} for the above-mentioned region of the western U.S.A.. 
   
     Figures 7 and 8 show the zonal wind spectra for troposphere and stratosphere respectively (the spectral data were taken from Fig. 6 of the Ref. \cite{nfg} and correspond to high wind speed cases). The dashed curves indicate the stretched exponential spectrum Eq. (15). The values of $\lambda_{\beta} = 2\pi/k_{\beta}$ indicate that the distributed chaos in the inertial range is tuned to the planetary waves both for the troposphere and stratosphere.

%%%%%%%%%%%%%%% 8 %%%%%%%%%%%%%%%%%%
\begin{figure} \vspace{-2cm}\centering
\epsfig{width=.46\textwidth,file=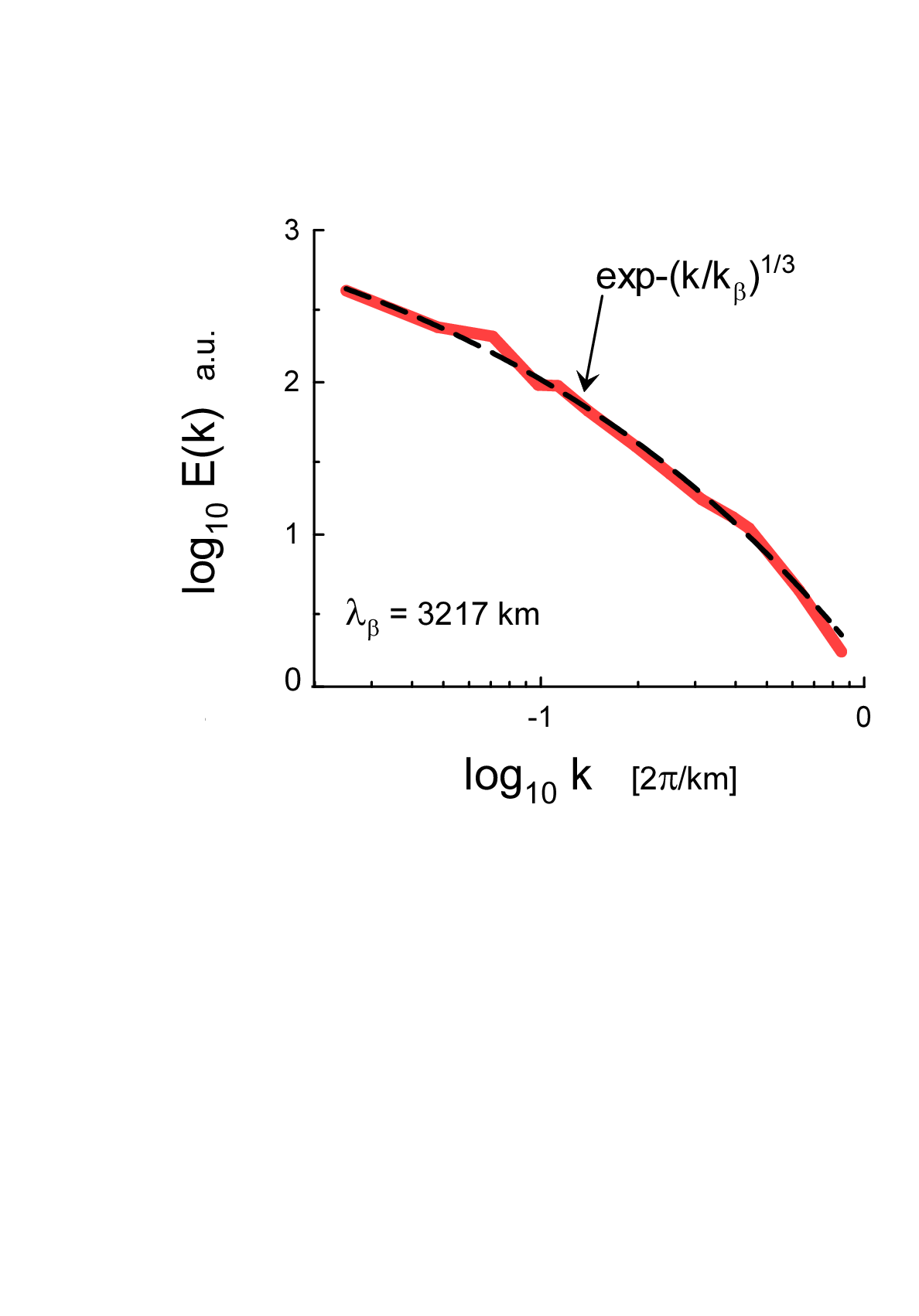} \vspace{-4.6cm}
\caption{As in Fig. 7 but for stratosphere.} 
\end{figure}
%%%%%%%%%%%%%%%%%%%%%%%%%%%%%%%%%%%%
%%%%%%%%%%%%%%% 9 %%%%%%%%%%%%%%%%%%
\begin{figure} \vspace{-0.4cm}\centering
\epsfig{width=.45\textwidth,file=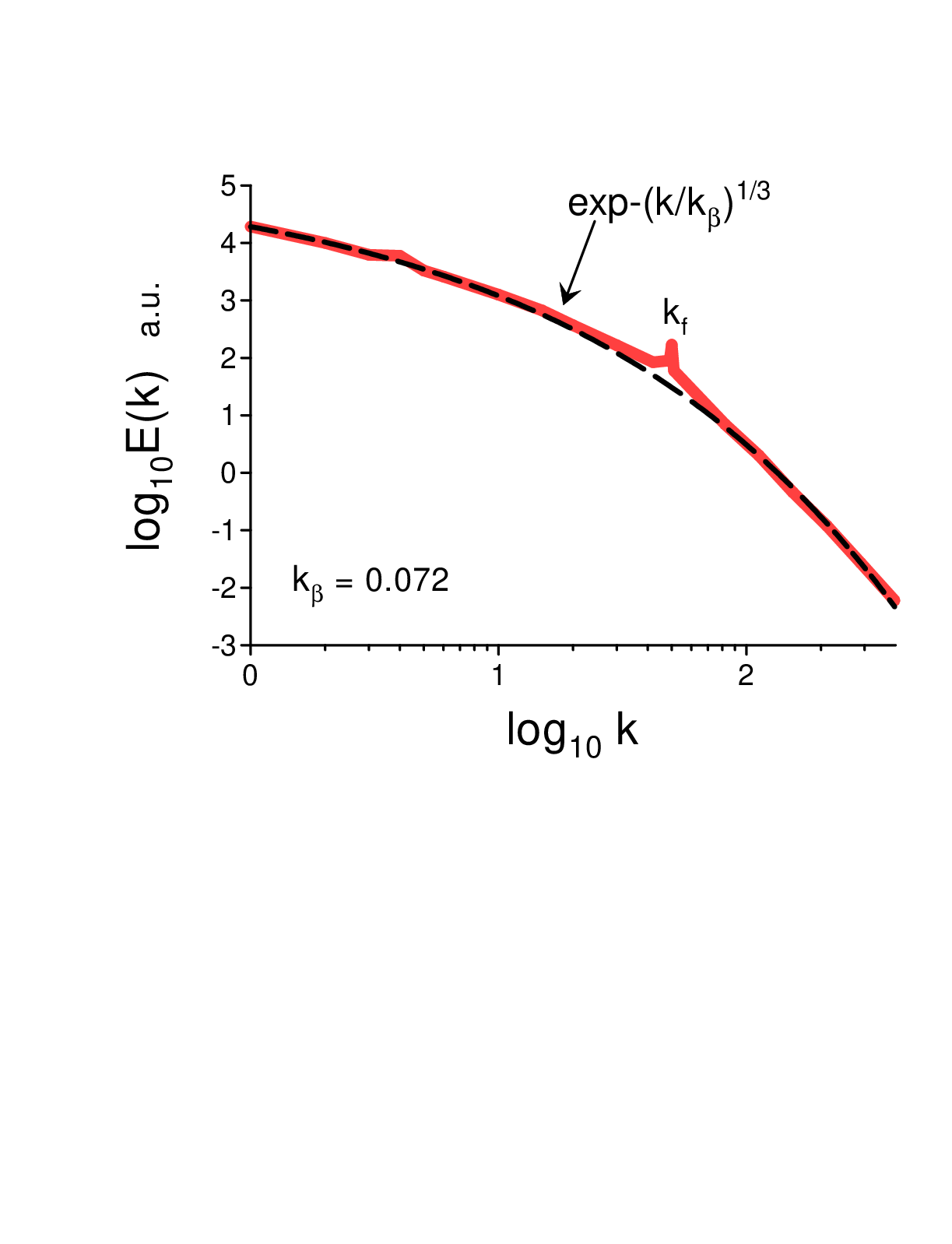} \vspace{-4.2cm}
\caption{The superfluid energy spectrum for the HVBK model.} 
\end{figure}
%%%%%%%%%%%%%%%%%%%%%%%%%%%%%%%%%%%%

\section{HVBK two-fluid model of superfluid}

A combination of the Euler (for the superfluid component) and the Navier-Stokes (for the normal component) equations related to each other by a mutual coupling term is known as the HVBK two-fluid model of the superfluid. This model is a classic-like one and can be applied for large spatial scales. \\

  In the paper Ref. \cite{sgp} results of a direct numerical simulation of an HVBK two-fluid model were reported. A specific HBVK system of equations was used in this DNS
$$
\frac{D {\bf v}_n}{D t} = -\frac{1}{\rho_n} \nabla p_n  - \eta_n {\bf v}_n + \frac{\rho_s}{\rho} {\bf F}_{ns} 
                    + \nu_n \nabla^2 {\bf v}_n  \eqno{(22)}
 $$
 $$                   
\frac{D {\bf v}_s}{D t} = -\frac{1}{\rho_s} \nabla p_s - \eta_s {\bf v}_s  - \frac{\rho_n}{\rho}{\bf F}_{ns} + \nu_s \nabla^2 {\bf v}_s
                    + {\bf f}_s^{ext},  \eqno{(23)}  
$$                            
and the incompressibility equations
$$
\nabla \cdot {\bf v}_n=0, ~~~~~\nabla \cdot {\bf v}_s=0     \eqno{(24-25)}   
$$
where the indexes $n$ and $s$ refer to the normal and the superfluid, $\rho=\rho_n+\rho_s$ is the total density.

  The mutual coupling term was taken as 
$$
{\bf F}_{ns} = -\frac{B}{2} \vert  {\boldsymbol \omega}_s \vert {\bf v}_{ns},  \eqno{(26)}
$$
where $B$ is a constant, ${\boldsymbol \omega}_s = [\nabla \times {\bf v_s}]$ is the superfluid vorticity, and ${\bf v}_{ns} = {\bf v}_n - {\bf v}_s$. 

The superfluid viscosity-like dissipation term $\nu_s \nabla^2 {\bf v}_s$ is suggested to take into account the small-scale effects that the large-scale HVBK model cannot directly take into account (such as the Kelvin waves and the quantized vortex reconnections \cite{kl},\cite{bpsl}\cite{vpk}). For the considered cases $\nu_n/\nu_s =10$ and $\rho_n /\rho = 0.1$ .

  To model the bottom or air-drag friction the  linear-friction terms $\eta_n {\bf v}_n$ and $\eta_s {\bf v}_s$ were used.

  The external superfluid forcing term ${\bf f}_s^{ext}$ was taken so that the superfluid vorticity field was forced by the term $ \propto \cos k_f x$ with $k_f =50$. 
  
    The HVBK equations were numerically solved in a square domain with periodic boundary conditions. \\
    
    Figure 9 shows the superfluid energy spectrum averaged over runs with $B =1,2$ and $5$ (the spectra for the individual values of $B = 1,2$ and $5$ are only slightly different). The spectral data were taken from Fig. 2b of the Ref. \cite{sgp}. The dashed curve indicates the stretched exponential spectrum Eq. (15).
    
 \section{Gross-Pitaevskii model for superfluid and  for Bose-Einstein condensate}  
 
 \subsection{Noether's theorem}
  
   When the normal fluid component is negligible one can use the Gross–Pitaevskii model as a good approximation for the superfluid $^4He$ and for the dilute Bose-Einstein condensate (BEC). 

  A simple dimensionless Gross–Pitaevskii equation is
$$
2i \frac{\partial \psi}{\partial t}=-\nabla^2 \psi +\left ( |\psi|^2-1\right )\psi, \eqno{(27)}
$$
where the coherence (or healing) length $\xi$ was taken for the unit of the spatial scale. The healing length $\xi$ is of the order of the quantized vortex core size. \\

    The Madelung's transformation $\psi = \sqrt{\rho} e^{i \theta}$ \cite{mad} allows to map Eq. (27) into classic like equation describing a compressible fluid \cite{bag},\cite{ya},\cite{sal}
$$
\rho \left(\frac{\partial v_i}{\partial t}+v_j\frac{\partial v_i}{x_j} \right) =
-\frac{\partial p}{\partial x_i} + \frac{\partial \tau_{ij}}{\partial x_j}  \eqno{(28)}
$$
where the density $\rho=|\psi|^2$, velocity ${\bf v}=\bm\nabla{\theta}$, $p=\rho^2/4$, and 
$$
\tau_{ij}=\frac{\rho}{4} \frac{\partial^2 \ln\rho}{\partial x_i \partial x_j} \eqno{(29)})
$$

    The $\tau_{ij}$ is a cause for the reconnections of the quantum vortices and is usually considered as quantum stress. At length scales much larger than the healing length $\xi$ the quantum stress term can be neglected in comparison to the dynamic pressure term and one obtains from Eq. (28) the usual Euler equation.\\

%%%%%%%%%%%%%%% 10 %%%%%%%%%%%%%%%%%%
\begin{figure} \vspace{-1.5cm}\centering
\epsfig{width=.48\textwidth,file=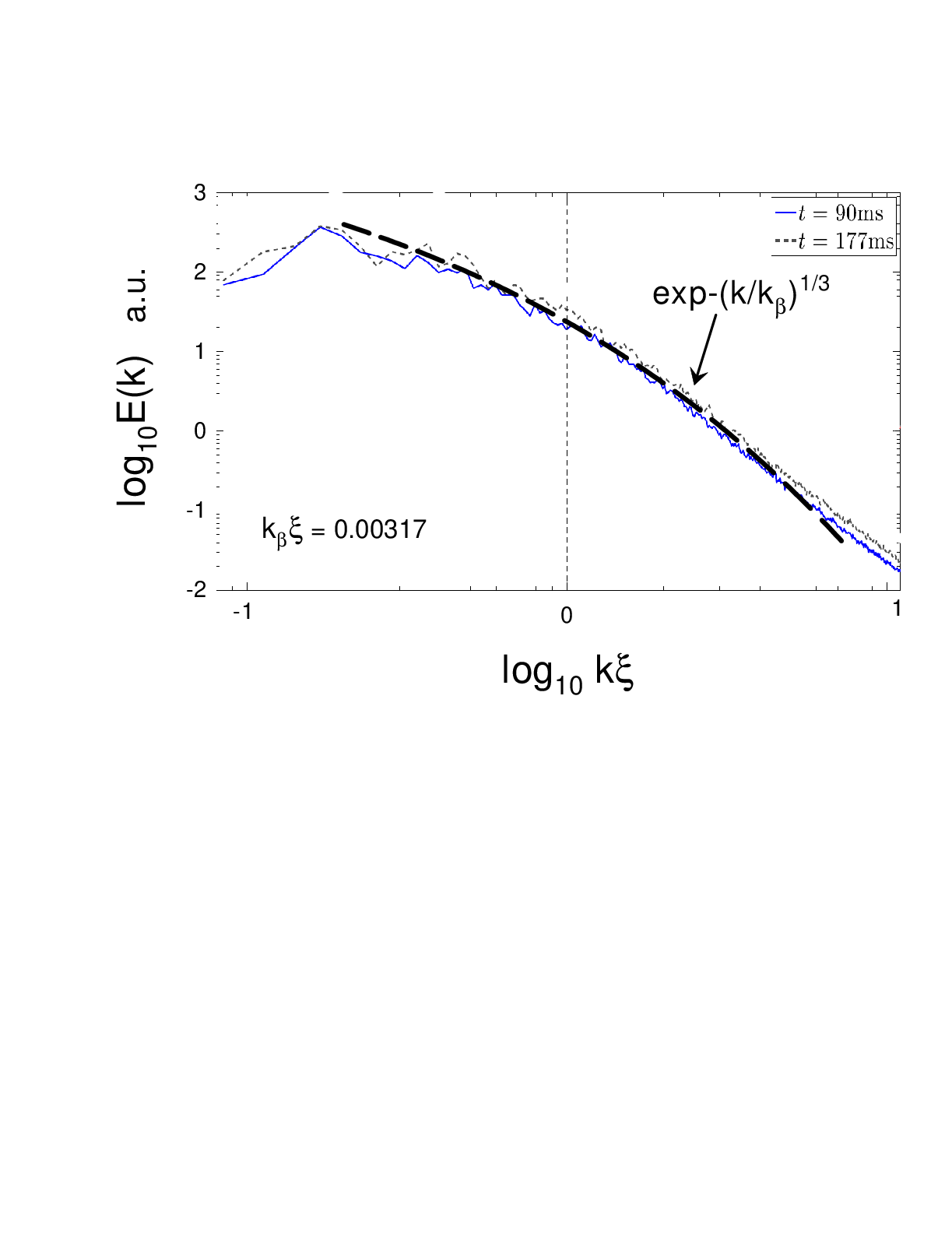} \vspace{-5.3cm}
\caption{ Incompressible energy spectrum for the Gross–Pitaevskii model: dipole regime.} 
\end{figure}
%%%%%%%%%%%%%%%%%%%%%%%%%%%%%%%%%%%%
%%%%%%%%%%%%%%% 11 %%%%%%%%%%%%%%%%%%
\begin{figure} \vspace{-0.4cm}\centering
\epsfig{width=.48\textwidth,file=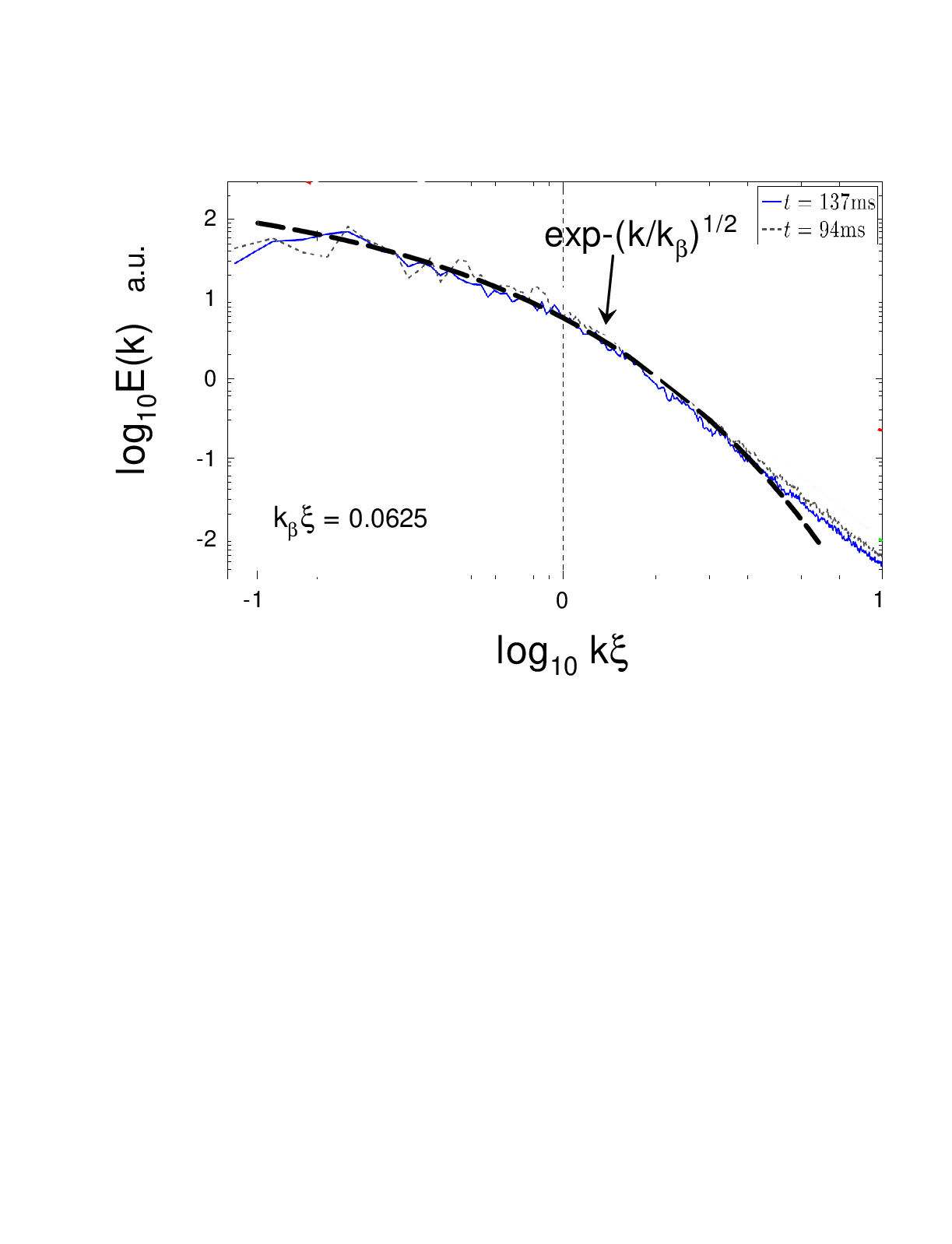} \vspace{-5.5cm}
\caption{ Incompressible energy spectrum for the Gross–Pitaevskii model: cluster regime} 
\end{figure}
%%%%%%%%%%%%%%%%%%%%%%%%%%%%%%%%%%%%
    
    It is known (see, for instance, Refs. \cite{fs},\cite{ls}) that the quasi-point vortex dynamical system is a good solution to the Gross–Pitaevskii equation in the asymptotic $\ell \gg \xi$ (where $\ell$ is the mean intervortex distance.). In this case, Noether's theorem provides conservation of the system's momentum and angular momentum \cite{ls} (cf Introduction). Therefore, the spectra Eqs. (15-16) can be applied to this case as well.
    
 \subsection{Direct numerical simulations}

    In paper Ref. \cite{rab} results of a direct numerical simulation were reported for the finite-temperature trapped Bose-Einstein condensate. The damped Gross-Pitaevskii equation was taken in the form 
$$
i\hbar \frac{\partial\psi({\bf r},t)}{\partial t} = L\psi({\bf r},t) + i\gamma\left[\mu - L\right]\psi({\bf r},t)  \eqno{(30)}
$$
where $\mu$ is the chemical potential and the operator
$$
L\psi({\bf r},t) \equiv \left[- \frac{\hbar^2\nabla_{\perp}^2}{2m} + V({\bf r},t) + g |\psi({\bf r},t)|^2 \right]\psi({\bf r},t)  \eqno{(31)}
$$
 In this simulation the phenomenological damping rate $\gamma$ was taken much smaller than all rates governing the system's dynamics.

   The external potential in Eq. (31) is a sum of the trapping potential $V_{tr}$ and the stirring potential $V_{st}$ 
$$
 V({\bf r},t) = V_{tr} + V_{st}   \eqno{(32)}
$$

  The system was initially three-dimensional but the cylindrically symmetric harmonic trapping potential was taken in the form
$$ 
V_{tr}({\bf r})==m\omega_r^2(x^2+y^2)/2+m\omega_z^2z^2/2  \eqno{(33)}
$$
with $\hbar \omega_z \gg \mu, k_BT, \hbar \omega_r$ ($r =\sqrt {(x^2+y^2})$ and $T$ is the temperature). Therefore, the system was effectively two-dimensional in the $(x,y)$ plane due to the strong confinement. 

The BEC superfluid was stirred by the two-dimensional repulsive potential
$$
V_s(x,y,t) = V_0 \exp \left[- \frac{(x - x_0(t))^2 + (y-y_0(t))^2}{\sigma^2}\right]  \eqno{(34)}
$$
with the stirring beam center located at the point $(x_0(t),y_0(t))$ and $\sigma = 4\xi$. This stirring potential can simulate a blue-detuned laser beam propagating along the axis $z$ (cf Section X). The circular motion of the potential barrier about the trap center was simulated by
$$
x_0 (t) = s\cos({\textrm v_{st}}t/s), ~~~ y_0 (t) = s \sin({\textrm v_{st}}t/s)   \eqno{(35)}
$$
with the stirrer speed $v_{st}$. \\

   The obstacle beam Eq. (34) is considered penetrable for $V_0 < \mu$ and impenetrable for $V_0 > \mu$. For $v_{st}/c \lesssim 0.3 $ (where $c$ is the sound velocity) the quantum vortex emission from the moving obstacle beam into the condensate does not occur. For $0.3   \lesssim v_{st}/c \lesssim 1$ the penetrable obstacle beam emitted about periodically the single quantum vortex dipoles (vortex-antivortex pairs) into the condensate. In this `dipole regime' there is no injection of the clusters of the vortices having the same circulation. But when the impenetrable obstacle was considered (at  $0.3   \lesssim v_{st}/c \lesssim 1$) irregular temporal emission of a mixture of quantum vortex dipoles and co-rotating quantum vortex clusters from the obstacle beam was observed  - the `cluster regime'. The temporal irregularity and clustering increased with increasing strength $V_0$ of the obstacle beam. It should be noted that the cross-over between the dipole and cluster regimes was gradual.\\

   Figures 10 and 11 show the incompressible energy spectra computed in this DNS for the dipole and cluster regimes respectively. The spectral data were taken from Figs. 3a and 3b of the Ref. \cite{rab}. The dashed curves indicate the stretched exponential spectra Eq. (15) and (16) for the dipole and cluster regimes respectively.
   
\section{Complex Ginzburg-Landau model}  

  The equation (30) can be considered as a particular case of the complex Ginzburg-Landau equation \cite{ara}. In a recent paper Ref. \cite{mit} results of a direct numerical simulation of the complex Ginzburg-Landau equation with a periodic drive were reported. The equation was taken in the form 
$$
 \frac{\partial A}{\partial t} = A+ (1+i \beta) \Delta A -(1+i \alpha) |A|^2 A +A_0 \delta (t- [T_0+l T])   \eqno{(37)}
$$
for the complex field $A ({\bf r}, t)$, $A_0$ is the strength of an external periodic driving field, $T_0$  represents the initial offset time, $T$ is the period of quenching, $l=0,1,2,..$. \\

  This equation allows solutions corresponding to topological coherent structures, the system of the vortices (in the Madelung's representation, see above), for instance.\\

  In this DNS the parameters were taken as $\alpha = 0.7$, $\beta = -0.7$, and $T =10$.  The simulation was initialized by $A = 0.001 \star R$ with $R$ uniformly distributed on the $[-0.5, 0.5]$, and an appropriate type of periodic boundary conditions was used. At the time $t = 1200$ (in the terms of the simulation) the system reached a state when the vortex-antivortex annihilation process was stopped (till $t = 2000$, at least), but the motion of the vortices was still observed (the so-called vortex glass regime). The state of the system at $t =1200$ was taken as the initial state for the amplitude quenches. Then the authors of the Ref. \cite{mit} periodically drove this system for different amplitudes $A_0$. \\ 
  
  Figures 12 shows the incompressible kinetic energy spectra for $t =1200 -2000$ with $A_0 =2$. The spectral data were taken from Fig. 7 of the Ref. \cite{mit}. The spectra for different values of $t$ were practically the same. The dashed curve indicates the stretched exponential spectrum Eq. (15). \\
  
  Figure 13  shows the incompressible kinetic energy spectrum for $t =2000$ with $A_0 =4.8$. The spectral data were taken from Fig. 7 of the Ref. \cite{mit}. One can see that still the spectrum can be well fitted by the stretched exponential Eq. (15), but now the parameters of the fitting curve are different from those corresponding to the case $A_0 =2$.

\section{A laboratory experiment} 

%%%%%%%%%%%%%%% 12 %%%%%%%%%%%%%%%%%%
\begin{figure} \vspace{-1.3cm}\centering
\epsfig{width=.45\textwidth,file=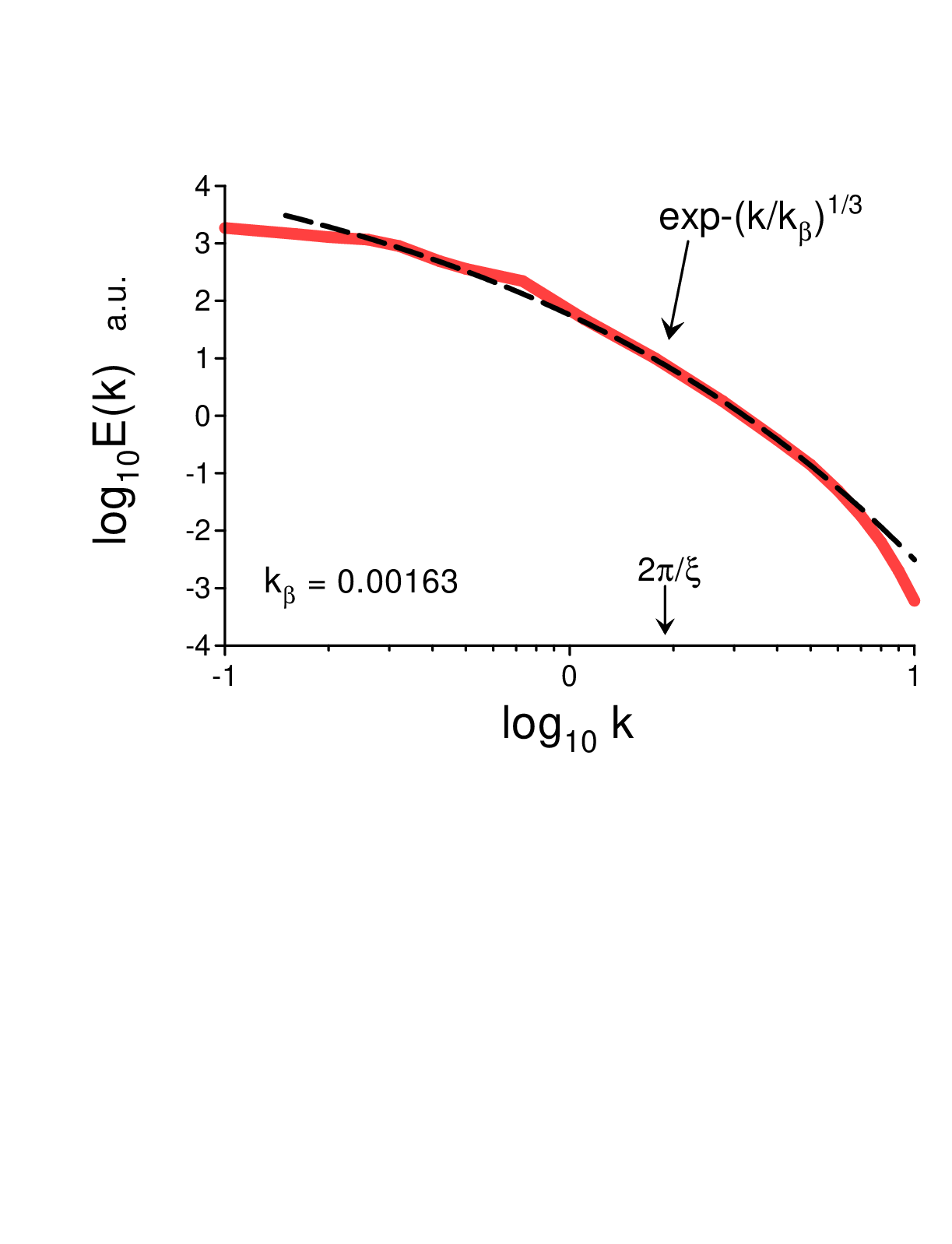} \vspace{-4.25cm}
\caption{Incompressible kinetic energy spectra for $t =1200 -2000$ with $A_0 =2$.} 
\end{figure}
%%%%%%%%%%%%%%%%%%%%%%%%%%%%%%%%%%%% 
%%%%%%%%%%%%%%% 13 %%%%%%%%%%%%%%%%%%
\begin{figure} \vspace{-0.5cm}\centering
\epsfig{width=.45\textwidth,file=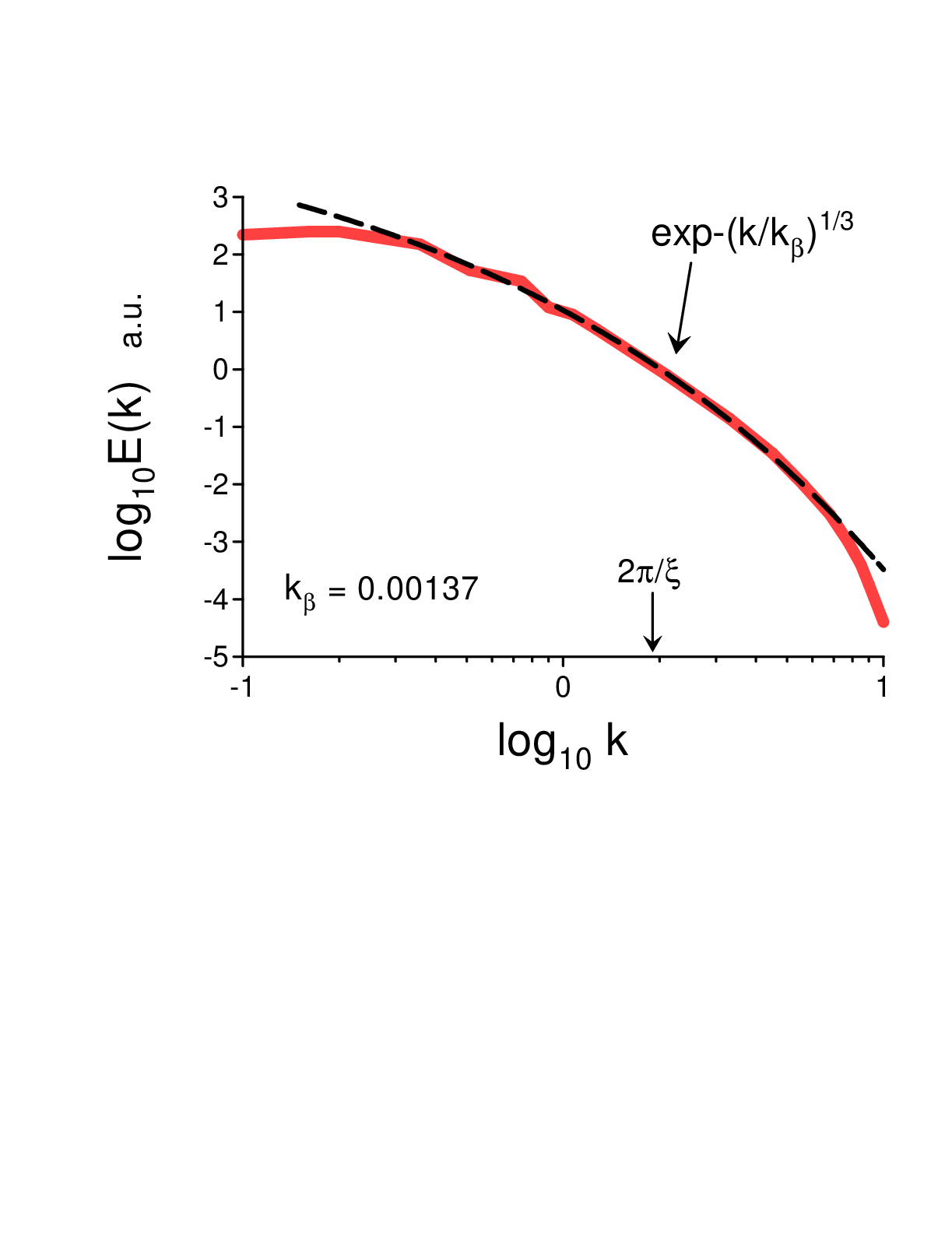} \vspace{-4.17cm}
\caption{Incompressible kinetic energy spectra for $t =2000$ with $A_0 =4.8$.} 
\end{figure}
%%%%%%%%%%%%%%%%%%%%%%%%%%%%%%%%%%%%

  In a recent paper Ref. \cite{john} results of a laboratory experiment with an oblate (quasi-two-dimensional) superfluid Bose-Einstein condensate were reported. The non-equilibrium distributions of vortices were generated by dragging a grid barrier through the condensate. The moving optical grid was formed by a moving array of laser beams. The velocity fluctuations in the BEC superfluid were measured by velocity-selective Bragg scattering. 
  
    Figure 14 shows the incompressible kinetic energy spectrum in the grid-generated BEC turbulence. The spectral data were taken from Fig. 4e of the Ref. \cite{john}. The system of vortices generated by the moving grid was dominated by the vortex dipole pairs in this case. The dashed curve indicates the spectrum Eq. (15). 
    
    It should be noted that by changing the parameters of the experiment (e.g. grid spacing) the authors of Ref. \cite{john} could change the character of the system of the grid injected vortices from the dipole to cluster one. But the spectral effect of this change was seen for the largest scales (comparable to the system size) only.

\section{Discussion}   

%%%%%%%%%%%%%%% 14 %%%%%%%%%%%%%%%%%%
\begin{figure} \vspace{-1.5cm}\centering
\epsfig{width=.48\textwidth,file=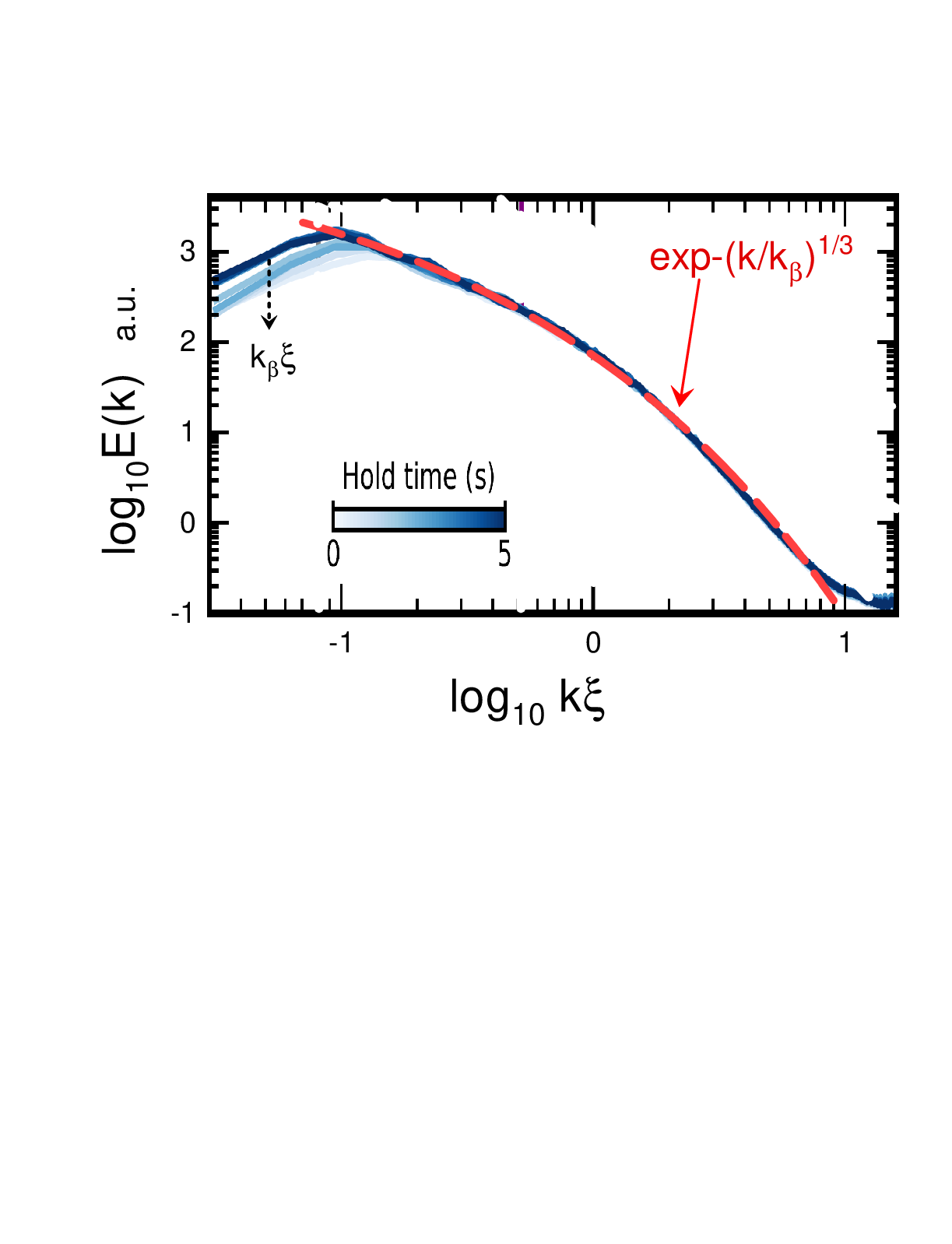} \vspace{-5.3cm}
\caption{ Incompressible kinetic energy spectrum in the grid-generated (dipole) laboratory BEC turbulence.} 
\end{figure}
%%%%%%%%%%%%%%%%%%%%%%%%%%%%%%%%%%%%

  It seems that isotropic homogeneous turbulence is more complex in two dimensions than in three dimensions. The two-dimensional turbulence is also more sensitive to the initial-boundary conditions and to the type of forcing (cf., for instance, recent review Ref. \cite{be} and references therein). This complexity and sensitivity can be related to the abundance of the invariants for the two-dimensional motion. The phenomenon of the quasi-point vortices (filaments, thin vortex tubes, quantum vortices) is characteristic also for three-dimensional turbulence under certain conditions (see, for instance, Refs. \cite{b1},\cite{yzs},\cite{els},\cite{isy},\cite{ked} and references therein) and should be taken into account in this case as well. \\

\section{Acknowledgement}

I thank J.C.R. Hunt and V. Yakhot for sending to me their papers.

\end{document}